\documentclass[prl,twocolumn]{revtex4}
\usepackage[pdftex]{graphicx}
\usepackage{amssymb}
\usepackage{textcomp}
\usepackage{latexsym}
\begin{document}
\pagestyle{plain}
\title{Phase-space for the breakdown of the quantum Hall effect in epitaxial graphene}
\author{J.A. Alexander-Webber$^1$, A.M.R. Baker$^1$, T.J.B.M. Janssen$^2$, A. Tzalenchuk$^2$, S. Lara-Avila$^3$, S. Kubatkin$^3$, R. Yakimova$^4$, B.A. Piot$^5$, D.K. Maude$^5$, and R.J. Nicholas$^1$}\email[Electronic address: ]{r.nicholas1@physics.ox.ac.uk}
\affiliation{$^1$Department of Physics, University of Oxford, Clarendon Laboratory, Parks Rd. Oxford OX1 3PU, U.K.}
\affiliation{$^2$National Physical Laboratory, Hampton Road, Teddington, TW11 0LW, U.K.}
\affiliation{$^3$Department of Microtechnology and Nanoscience, Chalmers University of Technology, S-412 96 G$\ddot{o}$teborg, Sweden}
\affiliation{$^4$Department of Physics, Chemistry and Biology (IFM), Link$\ddot{o}$ping University, S-581 83 Link$\ddot{o}$ping, Sweden}
\affiliation{$^5$LNCMI-CNRS-UJF-INSA-UPS, Grenoble, 38042 Cedex 9, France}

\date{\today}


\begin{abstract}
We report the phase-space defined by the quantum Hall effect breakdown in polymer gated epitaxial graphene on SiC (SiC/G) as a function of temperature, current, carrier density, and magnetic fields up to 30T. At 2K breakdown currents ($I_c$) almost two orders of magnitude greater than in GaAs devices are observed. The phase boundary of the dissipationless state ($\rho_{xx}=0$) shows a (1-$(T/T_c)^2$) dependence and persists up to $T_c>45K$ at 29T. With magnetic field $I_c$ was found to increase $\propto B^{3/2}$ and $T_c \propto B^{1.88}$. As the Fermi energy approaches the Dirac point, the $\nu=2$ quantized Hall plateau appears continuously from fields as low as 1T up to at least 19T due to a strong magnetic field dependence of the carrier density.
\end{abstract}
\maketitle

The quantum Hall effect (QHE) observed in two-dimensional electron gases is defined by vanishing longitudinal resistivity, $\rho_{xx}=0$, and a quantised Hall resistance, $\rho_{xy}=h/\nu e^2$ for $\nu=$integer. Ever since its first observation\cite{vklitzing:1980} in silicon the QHE has been used as a quantum electrical resistance standard which has been most extensively developed using GaAs devices\cite{Jeckelmann01}. In recent years since the the first isolation of graphene and the observation of the integer QHE \cite{novos05}\cite{zhang:2005} the attention of quantum Hall metrology labs has turned to graphene as potentially a more readily accessible resistance standard capable of operating at higher temperatures and measurement currents with lower magnetic fields. This is in part due to its large cyclotron energy gaps arising from the high electron velocity at the Dirac point. Recent experimental work\cite{Baker2012} has also shown that it has high electron-phonon energy relaxation rates, an order of magnitude faster than in GaAs heterostructures, which play an important role in determining the high current breakdown of the QHE. In particular polymer gated epitaxial graphene on SiC has been shown to be an exceptional candidate for metrology\cite{Tzalenchuk10,Janssen11}  and the universality of quantisation between it and GaAs has been shown to be accurate within a relative uncertainty of $8.6\times 10^{-11}$\cite{JanssenNJP11}.

If epitaxial graphene is to be used as a quantum resistance standard it is important to understand the experimental limits which confine the phase-space where the accurate, dissipationless QHE can be observed. Such a phase-space is determined by temperature, $T$, carrier density, $n$, magnetic field, $B$, and current $I$. The breakdown of the QHE is defined as the point where deviations from quantisation, $\Delta\rho_{xy}$ can be observed and this is strongly correlated with the point where $\rho_{xx}\neq 0$. A linear relationship of $\Delta\rho_{xy}/\rho_{xy}\propto\rho_{xx}$ is typically observed in GaAs\cite{Cage1984} and recently in graphene\cite{Guignard2012}, therefore measurement of the $I-V_{xx}$ characteristics in the quantum Hall regime also determines the maximum current consistent with maintaining a quantised $\rho_{xy}$. At high currents a sudden onset of longitudinal resistance is observed above a critical current $I_c$\cite{Jeckelmann01, Janssen11, Poirer09}. Modelling the breakdown using the bootstrap electron heating (BSEH) model from Komiyama and Kawaguchi \cite{Komiyama00} which looks at the balance between the rate of increase of energy input and electron energy loss has been reasonably successful. The Quantum Hall effect has already been reported in graphene at room temperature using magnetic fields of 45T\cite{Novoselov07}, however the plateaus did not show exact quantization as the resistivity was still finite ($\sim 10 \Omega$). In this work we address the formation of the zero-resistance state which corresponds to the dissipationless current flow needed to maintain the full quantum Hall condition.

Two devices were studied, prepared from epitaxially grown graphene on the Si-terminated face of SiC. Each device was lithographed using an e-beam and oxygen plasma etching into an eight leg Hall bar geometry ($W/L=4.5$) with widths of $W=5\mu$m and $W=35\mu$m for Sample 1 and Sample 2 respectively. Samples were electrically connected with large area Ti-Au contacting. A polymer gating technique using room temperature UV illumination was used to vary the electron density from $1-16\times 10^{11}$cm$^{-2}$ as described in Ref.\cite{Lara2011}. D.C. magnetotransport and $I-V$ data was taken using magnetic fields from a 21T superconducting solenoid and a 30T 20MW resistive-coil magnet at the LNCMI Grenoble.

\begin{figure}
\centering
\includegraphics[width=0.85\linewidth]{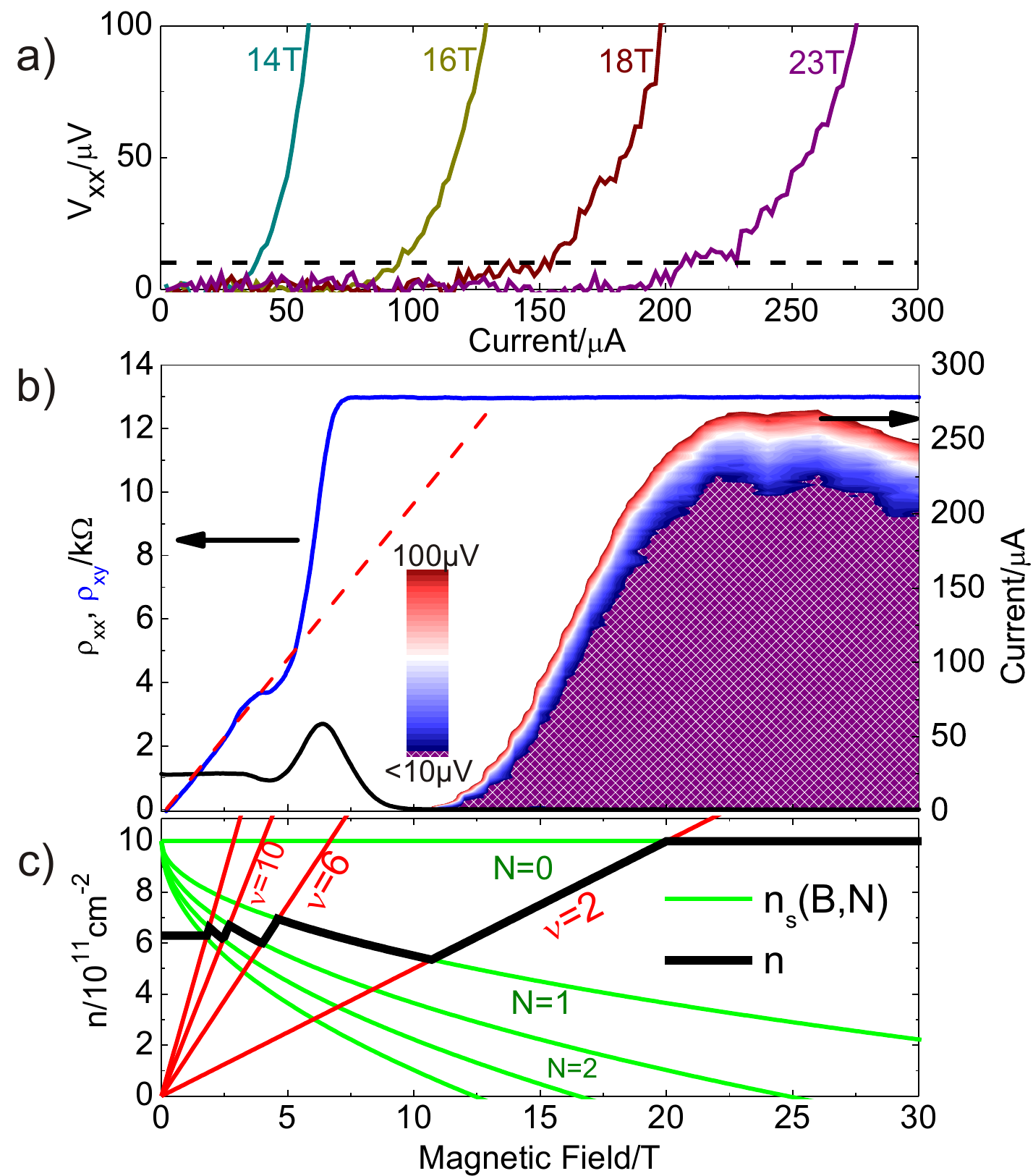}
\caption{a) $I-V_{xx}$ characteristics of Sample 1 at 2.0K, with a breakdown condition of $V_{xx}$ = 10$\mu$V, giving a maximum critical current density $j_c=43$A/m at 23T. b) Combined magneto-transport ($\rho_{xy}$(blue),$\rho_{xx}$(black)) data and $I-V_{xx}-B$ contour plot, the hashed region represents $V_{xx}<10\mu$V, the dissipationless quantum Hall regime. c) Magnetic field dependence of the carrier density (black), following lines of constant filling factor (red) while $E_F$ lies between Landau levels and then the charge transferred from surface donors in SiC, $n(B,N)$ (green), while the Landau levels fill, from the model in \cite{Janssen11}.}
\label{fig:Fig1}
\end{figure}

Fig. \ref{fig:Fig1}b shows $\rho_{xx}$ and $\rho_{xy}$ for Sample 1 with $n_{B=0}=6.5\times 10^{11}$cm$^{-2}$. We observe Shubnikov-de Haas oscillations in filling factors up to $\nu=8$, and a $\nu=2$ quantum Hall plateau beginning at $B=8$T with $\rho_{xx}=0$ from $B=10$T. This $\nu=2$ state is over 20T wide and observable all the way up to the maximum magnetic field of 30T. A series of $I-V_{xx}$ traces were taken every Tesla along the plateau to investigate the breakdown, with typical examples in Fig. \ref{fig:Fig1}a at $T=2K$. At 23T we find $V_{xx}=0$ until $I=I_c=215\mu$A, where we define the critical breakdown current at $V_{xx}(I_c)=10\mu$V, just above the noise-level of our measurements (Fig. \ref{fig:Fig1}a), corresponding to a resistivity of $\rho_{xx} \approx 0.01 \Omega$. Such a high breakdown current for a device just 5$\mu$m wide, giving a critical current density of $j_c=43A/m$, is truly exceptional in comparison to even the most well optimised GaAs devices ($j_c^{GaAs}\sim$1-2A/m). The full set of $I-V_{xx}$ traces are plotted in Fig. \ref{fig:Fig1}b as a contour plot. The hashed region is the the phase space where the dissipationless QHE is observed. The critical current $I_c$ increases along the plateau with a peak around 23T. Unlike traditional semiconductor quantum Hall systems which show a very sharp peak in $I_c$ centered at integer filling factor \cite{Alexander12}, the peak breakdown current occurs at fields much greater than $\nu=2$ calculated from the zero-field carrier density and changes very little in magnitude over a wide range of fields. This is due to the strong magnetic field dependence of the carrier density in epitaxial graphene grown on Si-terminated SiC \cite{Kopylov10}. Carriers are transferred to the graphene from the surface donor states of the SiC which are assumed to have a constant density of states. The charge transfer, $n_s(B,N)$ is proportional to the difference between the workfunction of the graphene and the SiC.  This causes the unbroadened Landau levels to be completely filled over a wide range of magnetic fields \cite{Janssen11}, particularly when the Fermi Energy $E_F$ is between the N=0 and N=1 Landau levels as in the region above 11T in Fig. \ref{fig:Fig1}c. Assuming that the peak $I_{c}$ occurs at $\nu=2$ suggests that the carrier density has increased to $n=1.1\times 10^{12}$cm$^{-2}$ by 23T and is still increasing. As a result the breakdown current is relatively independent of magnetic field which adds to the convenience of epitaxial graphene as an electrical resistance standard.

\begin{figure}
\centering
\includegraphics[width=0.85\linewidth]{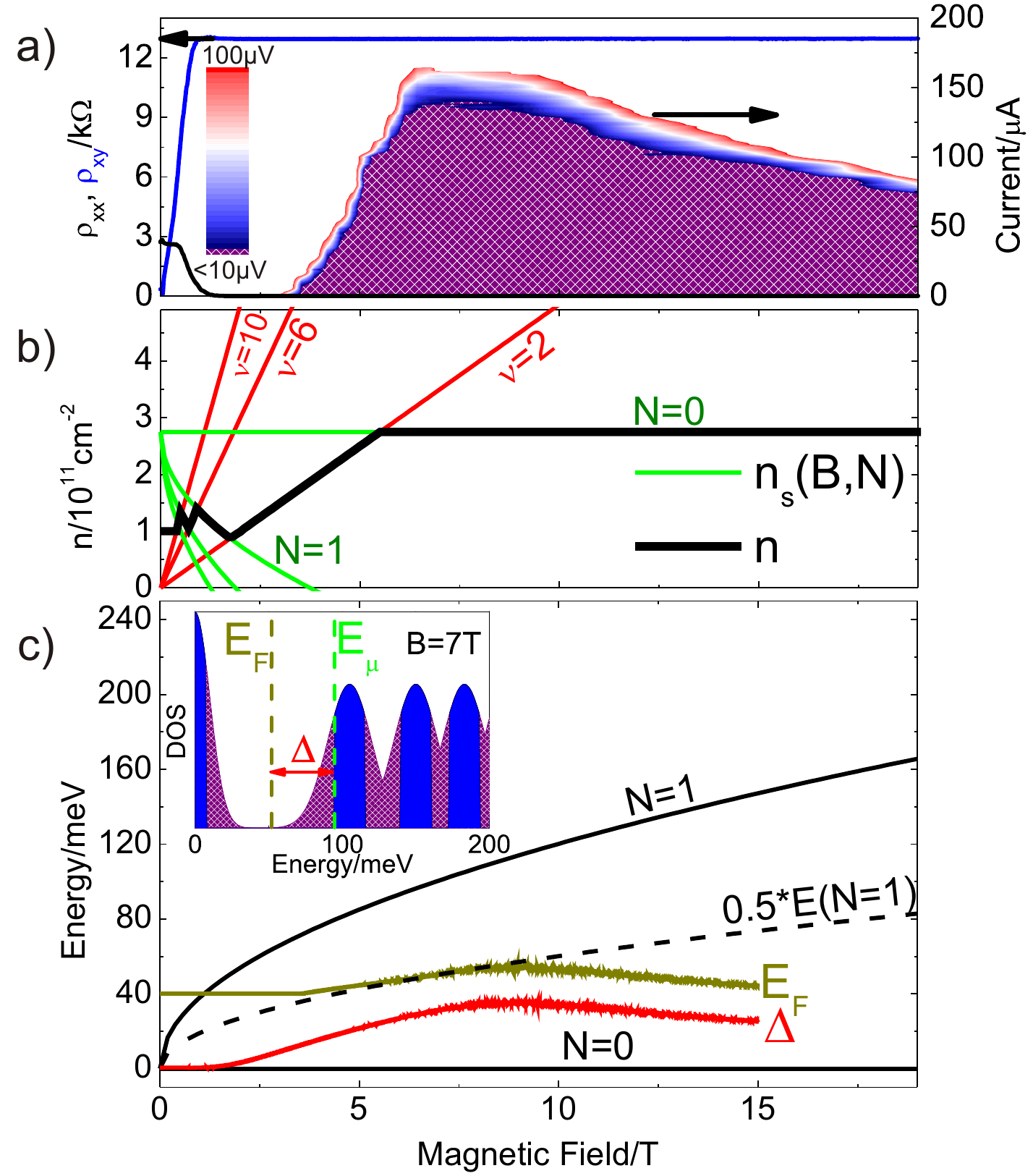}
\caption{a) Magnetotransport ($\rho_{xy}$(blue) and $\rho_{xx}$(black)) and corresponding $I-V_{xx}-B$ contour plot for the 35$\mu$m wide device at $T=$1.5K. b) Theoretical prediction of magnetic field dependent carrier density after Ref. \cite{Kopylov10}. c) $\Delta$ as a function of magnetic field and the resulting $E_F$.}
\label{fig:Fig2}
\end{figure}

At the lowest carrier density studied using Sample 2 ($n_{B=0}\sim 1\times 10^{11}$cm$^{-2}$), the $\nu=2$ state (Fig. \ref{fig:Fig2}a) begins at $B=$1T and persists up to the maximum field studied for this sample of 19T. The breakdown current shown in Fig. \ref{fig:Fig2}a is negligible at low fields ($B<$3T) but rapidly increases reaching a peak at $B=$7T, suggesting a carrier density of $n_{B=7T}=3.5\times 10^{11}$cm$^{-2}$. At 7T $I_c=140\mu$A, giving $j_c=4$A/m for this 35$\mu$m wide device. Importantly, from an applications perspective, $I_c\sim 100\mu$A by 5T, a magnetic field which is readily accessible with simple benchtop magnets. Applying the charge transfer model \cite{Kopylov10} the magnetic field for peak breakdown is accurately predicted (fig. \ref{fig:Fig2}b) but above this no further increase in carrier density is expected due to the finite density of donor states. The data suggest that the carrier density is still increasing, as the breakdown current has only decreased by a factor of 0.55 by 19T, probably due to the influence of level broadening which is not included in the original model\cite{Kopylov10}. In typical semiconductor 2DEGs \cite{Jeckelmann01, Alexander12}, breakdown currents show a triangular behaviour with a plateau width (defined by $I_c(\nu)/I_c\geqslant0$) of $\Delta\nu / \nu \sim \pm 0.2$. Assuming a level degeneracy ($\eta$) of 4 for the $\nu=2$ plateau due to the valley and spin degeneracies in graphene, this should correspond to a total plateau width of $\Delta\nu=\pm 0.8$, and $I_c$ should halve by 9T ($\nu=1.6$). This is consistent with results reported for exfoliated graphene \cite{Amado12}. By contrast the slow decrease in $I_c$ seen in Fig.\ref{fig:Fig2}a suggests that the occupancy remains $\nu \geqslant 1.6$ up to 19T where the carrier density has increased to $n \geqslant 7\times 10^{11}$cm$^{-2}$.

An Arrhenius analysis of the activated conductivity at higher temperatures (50-80K), above the variable range hopping regime\cite{Giesbers07, Bennaceur2012}, was used to estimate the magnetic field dependence of the Fermi energy, $E_F$, by measuring the activation gap $\Delta$ as a function of magnetic field. We assume that this measures the separation of $E_F$ from the conducting states $E_{\mu}$ of the nearest Landau level (N=1 for B$<$ 7T, N=0 for B$>$ 7T), where $\Delta=|E_{\mu}-E_F|$. Fig. \ref{fig:Fig2}c, shows $\Delta$ and the value of $E_F$ which has been deduced by assuming that it is midway between the two Landau levels at 7T where $\nu=2$.  At low fields $E_F$ corresponds to the approximately constant value of 40 meV deduced from the low field carrier density. Above 2.5T the carrier density begins to increase due to charge transfer from the substrate which keeps the Fermi energy in the gap between N=1 and N=0, and the system enters the dissipationless quantum Hall state. Above 7T $E_F$ falls slightly but appears pinned close to a constant energy of $E_F\sim 40$ meV suggesting that there may be a specific surface impurity level close to this value. This suggests that the $\nu=2$ plateau could extend up to higher fields still until the extended states of the symmetry broken N=0 state pass through the pinned Fermi level.

\begin{figure}
\centering
\includegraphics[width=0.85\linewidth]{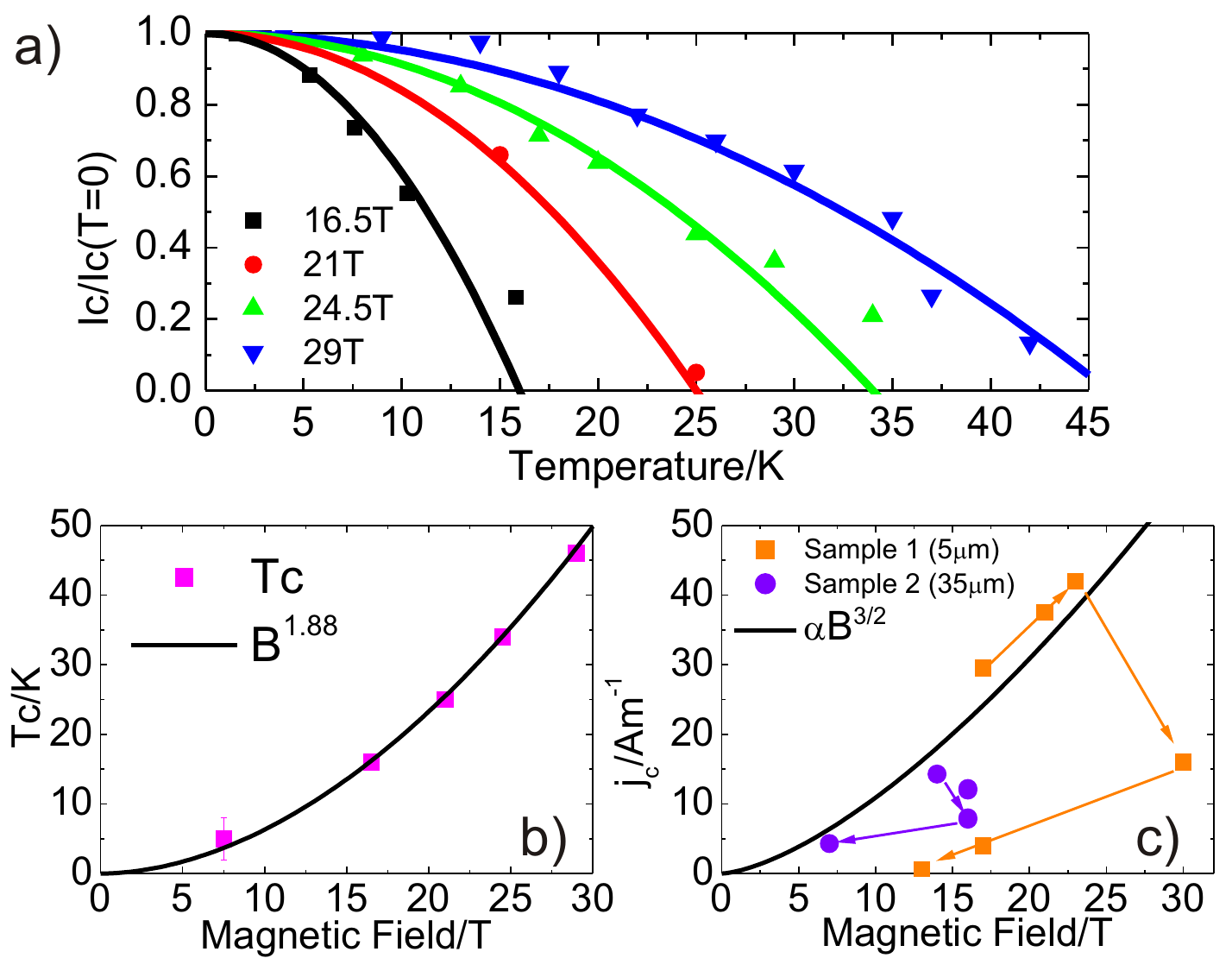}
\caption{a) Normalised temperature dependence of breakdown current at several magnetic fields, fitted with equation \ref{eq:temp}. (b) The magnetic field dependence of $T_c$, with a best fit of $0.08*B^{1.88}$. (c) The magnetic field dependence of $j_c$ for the two Hall bars with the sequence of illumination shown by arrows.}
\label{phase}
\end{figure}

In addition to the high currents observed above, high temperature ($T>4K$) operation is required for an accessible resistance standard. The temperature dependence of the breakdown was studied at several carrier densities for the peak $I_c$ at $\nu=2$, and for the highest carrier density of $n=1.6\times 10^{12}$cm$^{-2}$, at 29T as the maximum $I_c$ was just beyond our maximum field. In GaAs and InSb the temperature dependence \cite{Tanaka06, Rigal99, Alexander12} of the critical current has been shown to be of the form
\begin{equation}
I_c(T)=I_c(0) \left(1-\frac{T^2}{T_c^2}\right),
\label{eq:temp}
\end{equation}
where $T_c$ is the temperature at which $I_c=0$. Fig. \ref{phase}a shows that this equation also describes the temperature dependence of $I_c$ very well in epitaxial graphene for the magnetic fields studied. Rigal \textit{et al.} \cite{Rigal99} justified this based upon phenomenological similarities with the Gorter-Casimir two-fluid model for superconductors, and Tanaka \textit{et al.} \cite{Tanaka06} have proposed a model which predicts this behaviour based on a temperature-dependent mobility edge caused by the temperature dependence of the tunneling probabilites from localised to extended states at the centre of the Landau levels. Experimentally only limited evidence exists for the dependence of $T_c$ on magnetic field with values for GaAs\cite{Rigal99, Tanaka06} in the range 4.8 - 7.7K for B values of 4.8 - 7.7T at $\nu=4$ and $T_c \sim 1$/$\nu$, while for InSb $T_c$ = 8K at 6.1T\cite{Alexander12}, suggesting a linear dependence of $T_c$ upon field. It is therefore surprising that for graphene we see a strong superlinear scaling, as shown in Fig. \ref{phase}b with a best fit of $T_c\propto B^{1.88}$, which extrapolates to $T_c$ = 105K at 45T. The rate of increase of the the cyclotron energy gap between the $N$=1 and $N$=0 Landau level is sublinear, given by $E_N=sgn(N)\times c^{*}\sqrt{2e\hbar B|N|}$, where $c^{*}$ is the electron velocity, suggesting a weaker overall field dependence. One significant difference in epitaxial graphene is the magnitude of the disorder which means that the activation energy $\Delta=|E_{\mu}-E_F|$  at $\nu$=2 has a large offset due to level broadening and is known to increase more rapidly than the cyclotron energy\cite{Giesbers07} due possibly to smaller broadening for the N=0 Landau level which is topologically protected\cite{novos05}.

\begin{table}[t]
\centering
\begin{tabular} {c|c c c c c}
\hline\hline
Material & $\hbar\omega_c$ & $\tau_e$ & $j_c$ (A m$^{-1}$) & $j_c$ (A m$^{-1}$) & width\\
 & (meV) & (ps) & Theory\cite{Komiyama00} & Experiment & ($\mu$ m)\\
\hline
GaAs (7T) & 12 & 100\cite{Leadley1989} & 2.9 & 1.4 & 35\cite{Tanaka06}\\
InSb (7T) & 40 & 500\cite{Alexander12} & 2.6 & 0.3 & 600\cite{Alexander12}\\
Graphene (7T) & 105 & 80\cite{Baker2013} & 7.3 & \textbf{4.3} & 35\\
(17T) & 165 & 16\cite{Baker2013} & 36 & \textbf{30} & 5\\
(23T) & 200 & 6\cite{Baker2013} & 71 & \textbf{43} & 5\\
\hline
\end{tabular}
\caption{Material comparison for QHE breakdown at $\nu=2$}
\label{table:material}
\end{table}

By contrast $I_c$ has been extensively studied and is well known experimentally to scale as $B^{3/2}$\cite{Jeckelmann01,Poirer09,Tanaka06} as predicted by several of the models for breakdown\cite{Komiyama00} which include factors for the cyclotron energy and the inverse magnetic length. Fig. \ref{phase}c shows the values for $j_{c}$=$I_c$/W at $\nu=2$ for both samples after each UV illumination.  The highest values observed are also consistent with a $B^{3/2}$ dependence, although there is considerable spread in the experimental values, probably because the extended UV illumination is thought to introduce significant spatial inhomogeneities which are likely to reduce $j_c$. Interestingly despite the spread of $j_c$ values the same samples produced the very clear systematic dependence of $T_c$ shown in Fig. \ref{phase}. It should be noted that the values are significantly higher for the 5$\mu$m Hall bars and there is some evidence that quantum Hall breakdown current densities are larger for smaller Hall bar widths\cite{Jeckelmann01,Meziani2004}

The most widely accepted theory for the QHE breakdown is the bootstrap electron heating model proposed by Komiyama and Kawaguchi\cite{Komiyama00} in which the quantum Hall state becomes thermally unstable above a critical Hall electric field where the rate of change of electron-phonon energy loss rate becomes less than the rate of increase of input power. This predicts a critical breakdown electric field of
\begin{equation}
E_c = j_c\rho_{xy} = \sqrt{\frac{4B\hbar\omega_c}{\eta e\tau_e}},
\end{equation}
where $\tau_e$ is a characteristic electron-phonon energy relaxation time. Recently much experimental \cite{Baker2013, Betz2012, Tan11, Betz13} and theoretical\cite{Kubakaddi09} interest has focused on the way hot electrons lose energy to the lattice in graphene. We use the values of $\tau_e$, observed at $T_c$ from the damping of Shubnikov-deHaas oscillations\cite{Leadley89, Baker2013, Alexander12}, to calculate the predicted $I_c$ for $\nu=2$ and compare these to conventional semiconductor 2DEGs in Table \ref{table:material}. The graphene values are considerably larger, as compared for example to InSb, which has the lowest mass of the III-V semiconductors, $m^*=0.02m_e$\cite{Orr2009}. At 7T the cyclotron energy gap is 105meV for graphene, compared to 40meV in InSb, however we find an order of magnitude increase in current density for graphene over InSb. This is mainly a result of the factor 6 difference in $\tau_e$ between the two systems. The increase of $T_c$ with field causes $\tau_e$ to decrease and the dependence of $j_c$ on magnetic field to be superlinear. 

In summary, we have investigated the phase space in which the dissipationless quantum Hall state exists for epitaxial graphene. The data support the idea that this system can be described in terms of a phase diagram where the temperature dependence of the critical current follows a $\propto(1-(\frac{T}{T_c})^2)$ behaviour as seen in GaAs and InSb quantum Hall systems. We demonstrate that both the critical temperature and current are strongly magnetic field dependent and that at high fields critical current densities can be more than a factor 30 larger than previously observed in other systems. In epitaxial graphene, charge transfer from the carbon layer between the graphene and the SiC substrate also leads to a strongly magnetic field dependent carrier density and an exceptionally wide $\nu=2$ plateau due to charge transfer from surface impurities followed by pinning to a constant energy associated with a surface impurity level. 

\begin{acknowledgments}
This work was supported by EuroMagNET II, EU contract 228043, EU project ConceptGraphene, NPL Strategic Research Programme, and by
the UK EPSRC.
\end{acknowledgments}

%\bibliographystyle{apsrev}
%\bibliographystyle{unsrt}
%\bibliography{Generalbibliography2013}

\begin{thebibliography}{30}
\expandafter\ifx\csname natexlab\endcsname\relax\def\natexlab#1{#1}\fi
\expandafter\ifx\csname bibnamefont\endcsname\relax
  \def\bibnamefont#1{#1}\fi
\expandafter\ifx\csname bibfnamefont\endcsname\relax
  \def\bibfnamefont#1{#1}\fi
\expandafter\ifx\csname citenamefont\endcsname\relax
  \def\citenamefont#1{#1}\fi
\expandafter\ifx\csname url\endcsname\relax
  \def\url#1{\texttt{#1}}\fi
\expandafter\ifx\csname urlprefix\endcsname\relax\def\urlprefix{URL }\fi
\providecommand{\bibinfo}[2]{#2}
\providecommand{\eprint}[2][]{\url{#2}}

\bibitem[{\citenamefont{vonKlitzing et~al.}(1980)\citenamefont{vonKlitzing,
  Dorda, and Pepper}}]{vklitzing:1980}
\bibinfo{author}{\bibfnamefont{K.}~\bibnamefont{vonKlitzing}},
  \bibinfo{author}{\bibfnamefont{G.}~\bibnamefont{Dorda}}, \bibnamefont{and}
  \bibinfo{author}{\bibfnamefont{M.}~\bibnamefont{Pepper}},
  \bibinfo{journal}{Phys. Rev. Lett.} \textbf{\bibinfo{volume}{45}},
  \bibinfo{pages}{494} (\bibinfo{year}{1980}).

\bibitem[{\citenamefont{Jeckelmann and Jeanneret}(2001)}]{Jeckelmann01}
\bibinfo{author}{\bibfnamefont{B.}~\bibnamefont{Jeckelmann}} \bibnamefont{and}
  \bibinfo{author}{\bibfnamefont{B.}~\bibnamefont{Jeanneret}},
  \bibinfo{journal}{Rep. Prog. Phys.} \textbf{\bibinfo{volume}{64}},
  \bibinfo{pages}{1603} (\bibinfo{year}{2001}).

\bibitem[{\citenamefont{Novoselov et~al.}(2005)\citenamefont{Novoselov, Geim,
  Morozov, Jiang, Katsnelson, Grigorieva, Dubonos, and Firsov}}]{novos05}
\bibinfo{author}{\bibfnamefont{K.~S.} \bibnamefont{Novoselov}},
  \bibinfo{author}{\bibfnamefont{A.~K.} \bibnamefont{Geim}},
  \bibinfo{author}{\bibfnamefont{S.~V.} \bibnamefont{Morozov}},
  \bibinfo{author}{\bibfnamefont{D.}~\bibnamefont{Jiang}},
  \bibinfo{author}{\bibfnamefont{M.~I.} \bibnamefont{Katsnelson}},
  \bibinfo{author}{\bibfnamefont{I.~V.} \bibnamefont{Grigorieva}},
  \bibinfo{author}{\bibfnamefont{S.~V.} \bibnamefont{Dubonos}},
  \bibnamefont{and} \bibinfo{author}{\bibfnamefont{A.~A.}
  \bibnamefont{Firsov}}, \bibinfo{journal}{Nature}
  \textbf{\bibinfo{volume}{438}}, \bibinfo{pages}{197} (\bibinfo{year}{2005}).

\bibitem[{\citenamefont{Zhang et~al.}(2005)\citenamefont{Zhang, Tan, Stormer,
  and Kim}}]{zhang:2005}
\bibinfo{author}{\bibfnamefont{Y.}~\bibnamefont{Zhang}},
  \bibinfo{author}{\bibfnamefont{Y.~W.} \bibnamefont{Tan}},
  \bibinfo{author}{\bibfnamefont{H.~L.} \bibnamefont{Stormer}},
  \bibnamefont{and} \bibinfo{author}{\bibfnamefont{P.}~\bibnamefont{Kim}},
  \bibinfo{journal}{Nature} \textbf{\bibinfo{volume}{438}},
  \bibinfo{pages}{201} (\bibinfo{year}{2005}).

\bibitem[{\citenamefont{Baker et~al.}(2012)\citenamefont{Baker,
  Alexander-Webber, Altebaeumer, and Nicholas}}]{Baker2012}
\bibinfo{author}{\bibfnamefont{A.~M.~R.} \bibnamefont{Baker}},
  \bibinfo{author}{\bibfnamefont{J.~A.} \bibnamefont{Alexander-Webber}},
  \bibinfo{author}{\bibfnamefont{T.}~\bibnamefont{Altebaeumer}},
  \bibnamefont{and} \bibinfo{author}{\bibfnamefont{R.~J.}
  \bibnamefont{Nicholas}}, \bibinfo{journal}{Physical Review B}
  \textbf{\bibinfo{volume}{85}}, \bibinfo{pages}{115403}
  (\bibinfo{year}{2012}).

\bibitem[{\citenamefont{Tzalenchuk et~al.}(2010)\citenamefont{Tzalenchuk,
  Lara-Avila, Kalaboukhov, Paolillo, Syv{\"a}j{\"a}rvi, Yakimova, Kazakova,
  Janssen, Fal'ko, and Kubatkin}}]{Tzalenchuk10}
\bibinfo{author}{\bibfnamefont{A.}~\bibnamefont{Tzalenchuk}},
  \bibinfo{author}{\bibfnamefont{S.}~\bibnamefont{Lara-Avila}},
  \bibinfo{author}{\bibfnamefont{A.}~\bibnamefont{Kalaboukhov}},
  \bibinfo{author}{\bibfnamefont{S.}~\bibnamefont{Paolillo}},
  \bibinfo{author}{\bibfnamefont{M.}~\bibnamefont{Syv{\"a}j{\"a}rvi}},
  \bibinfo{author}{\bibfnamefont{R.}~\bibnamefont{Yakimova}},
  \bibinfo{author}{\bibfnamefont{O.}~\bibnamefont{Kazakova}},
  \bibinfo{author}{\bibfnamefont{T.~J. B.~M.} \bibnamefont{Janssen}},
  \bibinfo{author}{\bibfnamefont{V.}~\bibnamefont{Fal'ko}}, \bibnamefont{and}
  \bibinfo{author}{\bibfnamefont{S.}~\bibnamefont{Kubatkin}},
  \bibinfo{journal}{Nature Nanotech.} \textbf{\bibinfo{volume}{5}},
  \bibinfo{pages}{186} (\bibinfo{year}{2010}).

\bibitem[{\citenamefont{Janssen
  et~al.}(2011{\natexlab{a}})\citenamefont{Janssen, Tzalenchuk, Yakimova,
  Kubatkin, Lara-Avila, Kopylov, and Fal'ko}}]{Janssen11}
\bibinfo{author}{\bibfnamefont{T.~J. B.~M.} \bibnamefont{Janssen}},
  \bibinfo{author}{\bibfnamefont{A.}~\bibnamefont{Tzalenchuk}},
  \bibinfo{author}{\bibfnamefont{R.}~\bibnamefont{Yakimova}},
  \bibinfo{author}{\bibfnamefont{S.}~\bibnamefont{Kubatkin}},
  \bibinfo{author}{\bibfnamefont{S.}~\bibnamefont{Lara-Avila}},
  \bibinfo{author}{\bibfnamefont{S.}~\bibnamefont{Kopylov}}, \bibnamefont{and}
  \bibinfo{author}{\bibfnamefont{V.~I.} \bibnamefont{Fal'ko}},
  \bibinfo{journal}{Phys. Rev. B} \textbf{\bibinfo{volume}{83}},
  \bibinfo{pages}{233402} (\bibinfo{year}{2011}{\natexlab{a}}).

\bibitem[{\citenamefont{Janssen
  et~al.}(2011{\natexlab{b}})\citenamefont{Janssen, Fletcher, Goebel,
  Tzalenchuk, Yakimova, Kubatkin, Lara-Avila, and Fal'ko}}]{JanssenNJP11}
\bibinfo{author}{\bibfnamefont{T.~J. B.~M.} \bibnamefont{Janssen}},
  \bibinfo{author}{\bibfnamefont{N.~E.} \bibnamefont{Fletcher}},
  \bibinfo{author}{\bibfnamefont{J.~M.} \bibnamefont{Goebel},
  \bibfnamefont{R.~andWilliams}},
  \bibinfo{author}{\bibfnamefont{A.}~\bibnamefont{Tzalenchuk}},
  \bibinfo{author}{\bibfnamefont{R.}~\bibnamefont{Yakimova}},
  \bibinfo{author}{\bibfnamefont{S.}~\bibnamefont{Kubatkin}},
  \bibinfo{author}{\bibfnamefont{S.}~\bibnamefont{Lara-Avila}},
  \bibnamefont{and} \bibinfo{author}{\bibfnamefont{V.~I.}
  \bibnamefont{Fal'ko}}, \bibinfo{journal}{New J. Phys.}
  \textbf{\bibinfo{volume}{13}}, \bibinfo{pages}{093026}
  (\bibinfo{year}{2011}{\natexlab{b}}).

\bibitem[{\citenamefont{Cage et~al.}(1984)\citenamefont{Cage, Field, Dziuba,
  Girvin, Gossard, and Tsui}}]{Cage1984}
\bibinfo{author}{\bibfnamefont{M.}~\bibnamefont{Cage}},
  \bibinfo{author}{\bibfnamefont{B.}~\bibnamefont{Field}},
  \bibinfo{author}{\bibfnamefont{R.}~\bibnamefont{Dziuba}},
  \bibinfo{author}{\bibfnamefont{S.}~\bibnamefont{Girvin}},
  \bibinfo{author}{\bibfnamefont{A.}~\bibnamefont{Gossard}}, \bibnamefont{and}
  \bibinfo{author}{\bibfnamefont{D.}~\bibnamefont{Tsui}},
  \bibinfo{journal}{Physical Review B} \textbf{\bibinfo{volume}{30}},
  \bibinfo{pages}{2286} (\bibinfo{year}{1984}).

\bibitem[{\citenamefont{Guignard et~al.}(2012)\citenamefont{Guignard, Leprat,
  Glattli, Schopfer, and Poirier}}]{Guignard2012}
\bibinfo{author}{\bibfnamefont{J.}~\bibnamefont{Guignard}},
  \bibinfo{author}{\bibfnamefont{D.}~\bibnamefont{Leprat}},
  \bibinfo{author}{\bibfnamefont{D.~C.} \bibnamefont{Glattli}},
  \bibinfo{author}{\bibfnamefont{F.}~\bibnamefont{Schopfer}}, \bibnamefont{and}
  \bibinfo{author}{\bibfnamefont{W.}~\bibnamefont{Poirier}},
  \bibinfo{journal}{Physical Review B} \textbf{\bibinfo{volume}{85}},
  \bibinfo{pages}{165420} (\bibinfo{year}{2012}).

\bibitem[{\citenamefont{Poirer and Schopfer}(2009)}]{Poirer09}
\bibinfo{author}{\bibfnamefont{W.}~\bibnamefont{Poirer}} \bibnamefont{and}
  \bibinfo{author}{\bibfnamefont{F.}~\bibnamefont{Schopfer}},
  \bibinfo{journal}{Eur. Phys. J. Special Topics}
  \textbf{\bibinfo{volume}{172}}, \bibinfo{pages}{207} (\bibinfo{year}{2009}).

\bibitem[{\citenamefont{Komiyama and Kawaguchi}(2000)}]{Komiyama00}
\bibinfo{author}{\bibfnamefont{S.}~\bibnamefont{Komiyama}} \bibnamefont{and}
  \bibinfo{author}{\bibfnamefont{Y.}~\bibnamefont{Kawaguchi}},
  \bibinfo{journal}{Phys. Rev. B} \textbf{\bibinfo{volume}{61}},
  \bibinfo{pages}{2014} (\bibinfo{year}{2000}).

\bibitem[{\citenamefont{Novoselov et~al.}(2007)\citenamefont{Novoselov, Jiang,
  Zhang, Morozov, Stormer, Zeitler, Maan, Boebinger, Kim, and
  Geim}}]{Novoselov07}
\bibinfo{author}{\bibfnamefont{K.~S.} \bibnamefont{Novoselov}},
  \bibinfo{author}{\bibfnamefont{Z.}~\bibnamefont{Jiang}},
  \bibinfo{author}{\bibfnamefont{Y.}~\bibnamefont{Zhang}},
  \bibinfo{author}{\bibfnamefont{S.~V.} \bibnamefont{Morozov}},
  \bibinfo{author}{\bibfnamefont{H.~L.} \bibnamefont{Stormer}},
  \bibinfo{author}{\bibfnamefont{U.}~\bibnamefont{Zeitler}},
  \bibinfo{author}{\bibfnamefont{J.~C.} \bibnamefont{Maan}},
  \bibinfo{author}{\bibfnamefont{G.~S.} \bibnamefont{Boebinger}},
  \bibinfo{author}{\bibfnamefont{P.}~\bibnamefont{Kim}}, \bibnamefont{and}
  \bibinfo{author}{\bibfnamefont{A.~K.} \bibnamefont{Geim}},
  \bibinfo{journal}{Science} \textbf{\bibinfo{volume}{315}},
  \bibinfo{pages}{1379} (\bibinfo{year}{2007}).

\bibitem[{\citenamefont{Lara-Avila et~al.}(2011)\citenamefont{Lara-Avila,
  Moth-Poulsen, Yakimova, Bj\o~rnholm, Fal'ko, Tzalenchuk, and
  Kubatkin}}]{Lara2011}
\bibinfo{author}{\bibfnamefont{S.}~\bibnamefont{Lara-Avila}},
  \bibinfo{author}{\bibfnamefont{K.}~\bibnamefont{Moth-Poulsen}},
  \bibinfo{author}{\bibfnamefont{R.}~\bibnamefont{Yakimova}},
  \bibinfo{author}{\bibfnamefont{T.}~\bibnamefont{Bj\o~rnholm}},
  \bibinfo{author}{\bibfnamefont{V.}~\bibnamefont{Fal'ko}},
  \bibinfo{author}{\bibfnamefont{A.}~\bibnamefont{Tzalenchuk}},
  \bibnamefont{and} \bibinfo{author}{\bibfnamefont{S.}~\bibnamefont{Kubatkin}},
  \bibinfo{journal}{Adv. Mater.} \textbf{\bibinfo{volume}{23}},
  \bibinfo{pages}{878} (\bibinfo{year}{2011}).

\bibitem[{\citenamefont{Alexander-Webber
  et~al.}(2012)\citenamefont{Alexander-Webber, Baker, Buckle, Ashley, and
  Nicholas}}]{Alexander12}
\bibinfo{author}{\bibfnamefont{J.~A.} \bibnamefont{Alexander-Webber}},
  \bibinfo{author}{\bibfnamefont{A.~M.~R.} \bibnamefont{Baker}},
  \bibinfo{author}{\bibfnamefont{P.~D.} \bibnamefont{Buckle}},
  \bibinfo{author}{\bibfnamefont{T.}~\bibnamefont{Ashley}}, \bibnamefont{and}
  \bibinfo{author}{\bibfnamefont{R.~J.} \bibnamefont{Nicholas}},
  \bibinfo{journal}{Physical Review B} \textbf{\bibinfo{volume}{86}},
  \bibinfo{pages}{045404} (\bibinfo{year}{2012}).

\bibitem[{\citenamefont{Kopylov et~al.}(2010)\citenamefont{Kopylov, Tzalenchuk,
  Kubatkin, and Fal'ko}}]{Kopylov10}
\bibinfo{author}{\bibfnamefont{S.}~\bibnamefont{Kopylov}},
  \bibinfo{author}{\bibfnamefont{A.}~\bibnamefont{Tzalenchuk}},
  \bibinfo{author}{\bibfnamefont{S.}~\bibnamefont{Kubatkin}}, \bibnamefont{and}
  \bibinfo{author}{\bibfnamefont{V.~I.} \bibnamefont{Fal'ko}},
  \bibinfo{journal}{Appl. Phys. Lett.} \textbf{\bibinfo{volume}{97}},
  \bibinfo{pages}{112109} (\bibinfo{year}{2010}).

\bibitem[{\citenamefont{Amado et~al.}(2012)\citenamefont{Amado, Diez, Rossella,
  Bellani, Lopez-Romero, and Maude}}]{Amado12}
\bibinfo{author}{\bibfnamefont{M.}~\bibnamefont{Amado}},
  \bibinfo{author}{\bibfnamefont{E.}~\bibnamefont{Diez}},
  \bibinfo{author}{\bibfnamefont{F.}~\bibnamefont{Rossella}},
  \bibinfo{author}{\bibfnamefont{V.}~\bibnamefont{Bellani}},
  \bibinfo{author}{\bibfnamefont{D.}~\bibnamefont{Lopez-Romero}},
  \bibnamefont{and} \bibinfo{author}{\bibfnamefont{D.~K.} \bibnamefont{Maude}},
  \bibinfo{journal}{J. Phys: Condens. Matter} \textbf{\bibinfo{volume}{24}},
  \bibinfo{pages}{305302} (\bibinfo{year}{2012}).

\bibitem[{\citenamefont{Giesbers et~al.}(2007)\citenamefont{Giesbers, Zeitler,
  Katsnelson, Ponomarenko, Mohiuddin, and Maan}}]{Giesbers07}
\bibinfo{author}{\bibfnamefont{A.~J.~M.} \bibnamefont{Giesbers}},
  \bibinfo{author}{\bibfnamefont{U.}~\bibnamefont{Zeitler}},
  \bibinfo{author}{\bibfnamefont{M.~I.} \bibnamefont{Katsnelson}},
  \bibinfo{author}{\bibfnamefont{L.~A.} \bibnamefont{Ponomarenko}},
  \bibinfo{author}{\bibfnamefont{T.~M.} \bibnamefont{Mohiuddin}},
  \bibnamefont{and} \bibinfo{author}{\bibfnamefont{J.~C.} \bibnamefont{Maan}},
  \bibinfo{journal}{Phys. Rev. Lett.} \textbf{\bibinfo{volume}{99}},
  \bibinfo{pages}{206803} (\bibinfo{year}{2007}).

\bibitem[{\citenamefont{Bennaceur et~al.}(2012)\citenamefont{Bennaceur,
  Jacques, Portier, Roche, and Glattli}}]{Bennaceur2012}
\bibinfo{author}{\bibfnamefont{K.}~\bibnamefont{Bennaceur}},
  \bibinfo{author}{\bibfnamefont{P.}~\bibnamefont{Jacques}},
  \bibinfo{author}{\bibfnamefont{F.}~\bibnamefont{Portier}},
  \bibinfo{author}{\bibfnamefont{P.}~\bibnamefont{Roche}}, \bibnamefont{and}
  \bibinfo{author}{\bibfnamefont{D.~C.} \bibnamefont{Glattli}},
  \bibinfo{journal}{Phys. Rev. B} \textbf{\bibinfo{volume}{86}},
  \bibinfo{pages}{085433} (\bibinfo{year}{2012}).

\bibitem[{\citenamefont{Tanaka et~al.}(2006)\citenamefont{Tanaka, Kawashima,
  Iizuka, Fukuda, and Kawaji}}]{Tanaka06}
\bibinfo{author}{\bibfnamefont{H.}~\bibnamefont{Tanaka}},
  \bibinfo{author}{\bibfnamefont{H.}~\bibnamefont{Kawashima}},
  \bibinfo{author}{\bibfnamefont{H.}~\bibnamefont{Iizuka}},
  \bibinfo{author}{\bibfnamefont{H.}~\bibnamefont{Fukuda}}, \bibnamefont{and}
  \bibinfo{author}{\bibfnamefont{S.}~\bibnamefont{Kawaji}},
  \bibinfo{journal}{J. Phys. Soc. Japan} \textbf{\bibinfo{volume}{75}},
  \bibinfo{pages}{014701} (\bibinfo{year}{2006}).

\bibitem[{\citenamefont{Rigal et~al.}(1999)\citenamefont{Rigal, Maude,
  Potemski, Portal, Eaves, Wasilewski, Hill, and Pate}}]{Rigal99}
\bibinfo{author}{\bibfnamefont{L.~B.} \bibnamefont{Rigal}},
  \bibinfo{author}{\bibfnamefont{D.~K.} \bibnamefont{Maude}},
  \bibinfo{author}{\bibfnamefont{M.}~\bibnamefont{Potemski}},
  \bibinfo{author}{\bibfnamefont{J.~C.} \bibnamefont{Portal}},
  \bibinfo{author}{\bibfnamefont{L.}~\bibnamefont{Eaves}},
  \bibinfo{author}{\bibfnamefont{Z.~R.} \bibnamefont{Wasilewski}},
  \bibinfo{author}{\bibfnamefont{G.}~\bibnamefont{Hill}}, \bibnamefont{and}
  \bibinfo{author}{\bibfnamefont{M.~A.} \bibnamefont{Pate}},
  \bibinfo{journal}{Phys. Rev. Lett.} \textbf{\bibinfo{volume}{82}},
  \bibinfo{pages}{1249} (\bibinfo{year}{1999}).

\bibitem[{\citenamefont{Leadley
  et~al.}(1989{\natexlab{a}})\citenamefont{Leadley, Nicholas, Harris, and
  Foxon}}]{Leadley1989}
\bibinfo{author}{\bibfnamefont{D.~R.} \bibnamefont{Leadley}},
  \bibinfo{author}{\bibfnamefont{R.~J.} \bibnamefont{Nicholas}},
  \bibinfo{author}{\bibfnamefont{J.~J.} \bibnamefont{Harris}},
  \bibnamefont{and} \bibinfo{author}{\bibfnamefont{C.~T.} \bibnamefont{Foxon}},
  \bibinfo{journal}{Semicond. Sci. Technol.} \textbf{\bibinfo{volume}{4}},
  \bibinfo{pages}{879} (\bibinfo{year}{1989}{\natexlab{a}}).

\bibitem[{\citenamefont{Baker et~al.}(2013)\citenamefont{Baker,
  Alexander-Webber, Altebaeumer, McMullan, Janssen, Tzalenchuk, Lara-Avila,
  Kubatkin, Yakimova, Lin et~al.}}]{Baker2013}
\bibinfo{author}{\bibfnamefont{A.~M.~R.} \bibnamefont{Baker}},
  \bibinfo{author}{\bibfnamefont{J.~A.} \bibnamefont{Alexander-Webber}},
  \bibinfo{author}{\bibfnamefont{T.}~\bibnamefont{Altebaeumer}},
  \bibinfo{author}{\bibfnamefont{S.~D.} \bibnamefont{McMullan}},
  \bibinfo{author}{\bibfnamefont{T.~J. B.~M.} \bibnamefont{Janssen}},
  \bibinfo{author}{\bibfnamefont{A.}~\bibnamefont{Tzalenchuk}},
  \bibinfo{author}{\bibfnamefont{S.}~\bibnamefont{Lara-Avila}},
  \bibinfo{author}{\bibfnamefont{S.}~\bibnamefont{Kubatkin}},
  \bibinfo{author}{\bibfnamefont{R.}~\bibnamefont{Yakimova}},
  \bibinfo{author}{\bibfnamefont{C.-T.} \bibnamefont{Lin}},
  \bibnamefont{et~al.}, \bibinfo{journal}{Phys. Rev. B}
  \textbf{\bibinfo{volume}{87}}, \bibinfo{pages}{045414}
  (\bibinfo{year}{2013}).

\bibitem[{\citenamefont{Meziani et~al.}(2004)\citenamefont{Meziani, Chaubet,
  Bonifacie, Raymond, Poirier, and Piquemal}}]{Meziani2004}
\bibinfo{author}{\bibfnamefont{Y.~M.} \bibnamefont{Meziani}},
  \bibinfo{author}{\bibfnamefont{C.}~\bibnamefont{Chaubet}},
  \bibinfo{author}{\bibfnamefont{S.}~\bibnamefont{Bonifacie}},
  \bibinfo{author}{\bibfnamefont{A.}~\bibnamefont{Raymond}},
  \bibinfo{author}{\bibfnamefont{W.}~\bibnamefont{Poirier}}, \bibnamefont{and}
  \bibinfo{author}{\bibfnamefont{F.}~\bibnamefont{Piquemal}},
  \bibinfo{journal}{J. Appl. Phys.} \textbf{\bibinfo{volume}{96}},
  \bibinfo{pages}{404} (\bibinfo{year}{2004}).

\bibitem[{\citenamefont{Betz et~al.}(2012)\citenamefont{Betz, Vialla, Brunel,
  Voisin, Picher, Cavanna, Madouri, Feve, Berroir, Placais et~al.}}]{Betz2012}
\bibinfo{author}{\bibfnamefont{A.}~\bibnamefont{Betz}},
  \bibinfo{author}{\bibfnamefont{F.}~\bibnamefont{Vialla}},
  \bibinfo{author}{\bibfnamefont{D.}~\bibnamefont{Brunel}},
  \bibinfo{author}{\bibfnamefont{C.}~\bibnamefont{Voisin}},
  \bibinfo{author}{\bibfnamefont{M.}~\bibnamefont{Picher}},
  \bibinfo{author}{\bibfnamefont{A.}~\bibnamefont{Cavanna}},
  \bibinfo{author}{\bibfnamefont{A.}~\bibnamefont{Madouri}},
  \bibinfo{author}{\bibfnamefont{G.}~\bibnamefont{Feve}},
  \bibinfo{author}{\bibfnamefont{J.-M.} \bibnamefont{Berroir}},
  \bibinfo{author}{\bibfnamefont{B.}~\bibnamefont{Placais}},
  \bibnamefont{et~al.}, \bibinfo{journal}{Phys. Rev. Lett.}
  \textbf{\bibinfo{volume}{109}}, \bibinfo{pages}{056805}
  (\bibinfo{year}{2012}).

\bibitem[{\citenamefont{Tan et~al.}(2011)\citenamefont{Tan, Tan, Li, Liu, Lu,
  and Yang}}]{Tan11}
\bibinfo{author}{\bibfnamefont{Z.}~\bibnamefont{Tan}},
  \bibinfo{author}{\bibfnamefont{C.}~\bibnamefont{Tan}},
  \bibinfo{author}{\bibfnamefont{M.}~\bibnamefont{Li}},
  \bibinfo{author}{\bibfnamefont{G.~T.} \bibnamefont{Liu}},
  \bibinfo{author}{\bibfnamefont{L.}~\bibnamefont{Lu}}, \bibnamefont{and}
  \bibinfo{author}{\bibfnamefont{C.~L.} \bibnamefont{Yang}},
  \bibinfo{journal}{Phys. Rev. B} \textbf{\bibinfo{volume}{84}},
  \bibinfo{pages}{115429} (\bibinfo{year}{2011}).

\bibitem[{\citenamefont{Betz et~al.}(2013)\citenamefont{Betz, Jhang, Pallecchi,
  Ferreira, Fève, Berroir, and Plaçais}}]{Betz13}
\bibinfo{author}{\bibfnamefont{A.}~\bibnamefont{Betz}},
  \bibinfo{author}{\bibfnamefont{S.}~\bibnamefont{Jhang}},
  \bibinfo{author}{\bibfnamefont{E.}~\bibnamefont{Pallecchi}},
  \bibinfo{author}{\bibfnamefont{R.}~\bibnamefont{Ferreira}},
  \bibinfo{author}{\bibfnamefont{G.}~\bibnamefont{Fève}},
  \bibinfo{author}{\bibfnamefont{J.-M.} \bibnamefont{Berroir}},
  \bibnamefont{and} \bibinfo{author}{\bibfnamefont{B.}~\bibnamefont{Plaçais}},
  \bibinfo{journal}{Nature Physics} p. \bibinfo{pages}{109}
  (\bibinfo{year}{2013}).

\bibitem[{\citenamefont{Kubakaddi}(2009)}]{Kubakaddi09}
\bibinfo{author}{\bibfnamefont{S.~S.} \bibnamefont{Kubakaddi}},
  \bibinfo{journal}{Phys. Rev. B} \textbf{\bibinfo{volume}{79}},
  \bibinfo{pages}{075417} (\bibinfo{year}{2009}).

\bibitem[{\citenamefont{Leadley
  et~al.}(1989{\natexlab{b}})\citenamefont{Leadley, Nicholas, Harris, and
  Foxon}}]{Leadley89}
\bibinfo{author}{\bibfnamefont{D.~R.} \bibnamefont{Leadley}},
  \bibinfo{author}{\bibfnamefont{R.~J.} \bibnamefont{Nicholas}},
  \bibinfo{author}{\bibfnamefont{J.~J.} \bibnamefont{Harris}},
  \bibnamefont{and} \bibinfo{author}{\bibfnamefont{C.~T.} \bibnamefont{Foxon}},
  \bibinfo{journal}{Solid-State Electron.} \textbf{\bibinfo{volume}{32}},
  \bibinfo{pages}{1473} (\bibinfo{year}{1989}{\natexlab{b}}).

\bibitem[{\citenamefont{Orr et~al.}(2009)\citenamefont{Orr, Chuang, Nicholas,
  Buckle, Emeny, and Buckle}}]{Orr2009}
\bibinfo{author}{\bibfnamefont{J.~M.~S.} \bibnamefont{Orr}},
  \bibinfo{author}{\bibfnamefont{K.~C.} \bibnamefont{Chuang}},
  \bibinfo{author}{\bibfnamefont{R.~J.} \bibnamefont{Nicholas}},
  \bibinfo{author}{\bibfnamefont{L.}~\bibnamefont{Buckle}},
  \bibinfo{author}{\bibfnamefont{M.~T.} \bibnamefont{Emeny}}, \bibnamefont{and}
  \bibinfo{author}{\bibfnamefont{P.~D.} \bibnamefont{Buckle}},
  \bibinfo{journal}{Phys. Rev. B} \textbf{\bibinfo{volume}{79}},
  \bibinfo{pages}{235302} (\bibinfo{year}{2009}).

\end{thebibliography}

\end{document}